\documentclass[aip,jmp,a4paper,reprint,groupedaddress]{revtex4-1}
\usepackage{amsmath,amssymb,amsthm,mathtools,graphicx}

\newcommand{\ir}{\mathrm{i}}
\newcommand{\N}{\mathbb{N}}
\newcommand{\F}{\mathbb{F}}
\newcommand{\cA}{\mathcal{A}}
\newcommand{\cB}{\mathcal{B}}
\newcommand{\cH}{\mathcal{H}}
\newcommand{\one}{\openone}

\newcommand{\bra}[1]{\langle #1\vert}
\newcommand{\ket}[1]{\vert #1\rangle}
\newcommand{\braket}[2]{\langle #1\vert #2\rangle}
\newcommand{\ketbra}[1]{\vert #1\rangle\langle #1\vert}
\newcommand{\ev}[1]{\langle #1\rangle}
\newcommand{\abs}[1]{\vert #1\vert}
\newcommand{\bigabs}[1]{\bigl\vert #1\bigr\vert}
\newcommand{\comm}[2]{[#1,#2]}
\newcommand{\anticomm}[2]{\{#1,#2\}}

\DeclareMathOperator{\diag}{diag}
\DeclareMathOperator{\tr}{Tr}

\newtheorem*{MU}{Maassen--Uffink uncertainty relation}
\newtheorem*{marriage}{Marriage theorem}
\newtheorem{theorem}{Theorem}
\newtheorem{lemma}[theorem]{Lemma}

\begin{document}

\title{Entropic uncertainty relations and the stabilizer formalism}
\date{2 February 2012}
\author{S\"onke Niekamp}
\author{Matthias Kleinmann}
\author{Otfried G\"uhne}
\affiliation{Universit\"at Siegen, Fachbereich Physik,
Walter-Flex-Stra\ss{}e 3, D-57068 Siegen, Germany}
\affiliation{Institut f\"ur Quantenoptik und Quanteninformation,
\"Osterreichische Akademie der Wissenschaften,
Technikerstra\ss{}e 21a, A-6020 Innsbruck, Austria}

\begin{abstract}
  Entropic uncertainty relations express the quantum mechanical uncertainty
  principle by quantifying uncertainty in terms of entropy. Central questions
  include the derivation of lower bounds on the total uncertainty for given
  observables, the characterization of observables that allow strong
  uncertainty relations, and the construction of such relations for the case of
  several observables. We demonstrate how the stabilizer formalism can be
  applied to these questions. We show that the Maassen--Uffink entropic
  uncertainty relation is tight for the measurement in any pair of stabilizer
  bases. We compare the relative strengths of variance-based and various
  entropic uncertainty relations for dichotomic anticommuting observables.
\end{abstract}

\maketitle

\section{Introduction}

In quantum mechanics, one cannot predict a measurement outcome with certainty
unless the system is in an eigenstate of the observable being measured. It
follows that if two or more observables have no common eigenstate, it is not
possible to prepare the system such that for each observable only one
measurement outcome can occur. This is known as the uncertainty
principle,~\cite{Heis27,Kenn27,Robe29,BuHL07} which is quantitatively
formulated in terms of uncertainty relations.

Uncertainty relations not only describe a fundamental quantum mechanical
concept, but have also found application, e.\,g., in quantum
cryptography~\cite{Koas06,DFRS07,Koas09} and entanglement
detection.~\cite{HoTa03,GuLe04,Gueh04b} Entropic uncertainty
relations~\cite{BBMy75,Deut83,Krau87,MaUf88,WeWi10} have turned out to be
particularly useful. More recently, the uncertainty principle has also been
formulated in terms of majorization relations.~\cite{Part11} In
Ref.~\onlinecite{BCCR10}, Berta \emph{et al.} derived an entropic uncertainty
relation for a system which is entangled to a quantum memory. Access to this
memory can then be used to lower the uncertainties of measurements on the
system. This relation has been the subject of recent
experiments.~\cite{PHCF11,LXXL11}

Historically, the first and still the most celebrated uncertainty relation was
given by Heisenberg~\cite{Heis27,Kenn27} and applies to canonically conjugate
observables such as position and momentum, stating that
$\Delta^2(q)\Delta^2(p)\ge\hbar^2/4$. It was generalized to arbitrary
observables in the form of Robertson's~\cite{Robe29} uncertainty relation
$\Delta^2(A)\Delta^2(B)\ge\abs{\ev{\comm{A}{B}}}^2/4$. On closer inspection,
the latter does not have all desirable properties of an uncertainty relation,
in particular since it is trivial for any eigenstate of either observable.
Robertson's relation has also been generalized to the case of more than two
observables.~\cite{Robe34,Trif02}

A different approach is based on the idea of quantifying uncertainty by the
entropy of the probability distribution for the measurement
outcomes.~\cite{BBMy75,Deut83,Krau87,MaUf88,WeWi10} As a consequence, the
resulting uncertainty relations depend only on the eigenstates, but not on the
eigenvalues of the observables. That is, they are independent of the labelling
of the measurement results, which in finite dimensions is essentially
arbitrary. For this reason, entropic uncertainty relations can be regarded as a
more natural formulation of the uncertainty principle, at least in the case of
finitely many measurement outcomes, which we consider here. For a review of
entropic uncertainty relations see, e.\,g., Ref.~\onlinecite{WeWi10}.

An entropic uncertainty relation [see Eq.~\eqref{eq:mu} for an example] gives a
lower bound on the sum of the entropies for the measurement outcomes of some
observables on a quantum state. This bound necessarily depends on the
observables, but preferably is independent of the state. For the case of two
observables, such bounds have been determined, e.\,g., in
Refs.~\onlinecite{Deut83,Krau87,MaUf88}. Finding the best bound is in general
not easy. Putting differently, the question here is a characterization of
observables which admit strong uncertainty relations. So far, most results
apply to the case of two observables only,~\cite{WeWi08,WeWi10} despite the
generalization of the theory to several observables being an interesting
problem.

In this article, we demonstrate how the stabilizer formalism~\cite{HDER06} can
be applied to these questions. This formalism provides an efficient description
of certain many-qubit states, including highly entangled ones. In the field of
quantum information theory, stabilizer stabilizer states and the more special
case of graph states are widely used. For example, a particular class of graph
states, namely, cluster states, serve as a universal resource for
measurement-based quantum computation.~\cite{RaBr01,RaBB03} Cluster states are
also interesting because they are particularly robust against decoherence,
independently of the system size.~\cite{DuBr04,HeDB05} Graph states have been
employed as codewords of error-correcting codes.~\cite{ScWe01} Finally, the
stabilizer formalism has been used for entanglement detection.~\cite{ToGu05}

Using the stabilizer formalism, we first investigate the question for which
observables the Maassen--Uffink uncertainty relation [see Eq.~\eqref{eq:mu}] is
tight. We show that this is the case for the measurement in any two stabilizer
bases. We then turn our attention to the many-observable setting, focussing on
dichotomic anticommuting observables. Generalizing a result by Wehner and
Winter,~\cite{WeWi08} we provide a systematic construction of uncertainty
relations. The family of uncertainty relations we obtain contains both entropic
and variance-based ones. We compare the relative strengths of these relations.
Finally, we apply them to the stabilizing operators of two stabilizer states.

This article is organized as follows: In Sec.~\ref{sec:statement}, we introduce
our notation and review previous results. This includes a short introduction to
stabilizers and graph states. Section~\ref{sec:strong} contains our results on
measurements in stabilizer bases, and Sec.~\ref{sec:anticommuting} those on
several dichotomic anticommuting observables. In Sec.~\ref{sec:stabilizer}, the
latter are applied to the stabilizing operators of two stabilizer states.
Finally, Sec.~\ref{sec:conclusion} is devoted to a discussion of the results.

\section{Statement of the problem}
\label{sec:statement}

\subsection{Entropic uncertainty relations}

We consider the following general situation: when measuring an observable $A$
on a state $\rho$, the measurement outcome $a_i$ occurs with probability
$p_i=\tr(\Pi_i\rho)$, where $A=\sum_{i=1}^ma_i\Pi_i$ is the spectral
decomposition of the observable and the $a_i$ are mutually distinct. For an
entropy function $S$, we denote by $S(A\vert\rho)$ the entropy of this
probability distribution $P=(p_1,\ldots,p_m)$. In this article, we will use the
Shannon entropy
\begin{equation}
  S^S(P)=-\sum_{i=1}^{m}p_i\log(p_i)
\end{equation}
as well as the min-entropy
\begin{equation}
  S^\text{min}(P)=-\log(\max_i p_i)
\end{equation}
and the Tsallis entropy
\begin{equation}
  S_q^T(P)=\frac{1-\sum_{i=1}^m(p_i)^q}{q-1},\qquad q>1.
\end{equation}
In the limit $q\to1$, the Tsallis entropy gives the Shannon entropy. We use the
logarithm to base $2$ throughout.

We are interested in uncertainty relations for a family of observables
$\{A_1,\ldots,A_L\}$ of the form
\begin{equation}
  \label{eq:ur}
  \frac{1}{L}\sum_{k=1}^LS(A_k\vert\rho)\ge c_{\{A_k\}},
\end{equation}
where $S$ is an entropy function. The lower bound $c_{\{A_k\}}$ may depend on
the observables, but shall be independent of the state. For a given set of
observables, an uncertainty relation is called tight, if a state $\rho_0$
exists that attains the lower bound,
$1/L\sum_{k=1}^LS(A_k\vert\rho_0)=c_{\{A_k\}}$.

Clearly the entropy of an observable depends only on its eigenstates and is
independent of its eigenvalues, as long as they are nondegenerate. We will
therefore not distinguish between a nondegenerate observable $A$ and its
eigenbasis $\cA$. The Shannon entropy satisfies $0\le
S^S(\cA\vert\rho)\le\log(m)$, where $m$ is the length of the basis $\cA$. If we
choose for $\rho$ one of the basis states of $\cA_k$, we have
$1/L\sum_{k=1}^LS(\cA_k\vert\rho)\le\log(m)(L-1)/L$, because in this case the
entropy is zero for one basis and upper bounded by $\log(m)$ for the remaining
$L-1$ bases. This implies that for the Shannon entropy the right-hand side of
Eq.~\eqref{eq:ur} cannot exceed $\log(m)(L-1)/L$. An uncertainty relation that
reaches this limit is called maximally strong, and the corresponding
measurements are called maximally incompatible.~\cite{WeWi10} In other words,
maximal incompatibility means that if the outcome of one measurement is
certain, the outcomes of the remaining measurements are completely random.

A related notion is mutual unbiasedness. Two orthonormal bases $\ket{a_i}$ and
$\ket{b_i}$, $i=1,\ldots,d$, are called mutually unbiased if
\begin{equation}
  \bigabs{\braket{a_i}{b_j}}=\frac{1}{\sqrt{d}},
\end{equation}
for all $i$ and $j$. Pairwise mutual unbiasedness is a necessary, but for more
than two bases not a sufficient condition for the existence of a maximally
strong uncertainty relation.~\cite{WeWi10}

For the Shannon entropies of $L=2$ measurement bases, Maassen and
Uffink~\cite{MaUf88} have proven the following result:

\begin{MU}
  For any two measurement bases $\cA=\{\ket{a_i}\}$ and $\cB=\{\ket{b_i}\}$,
  \begin{equation}
    \label{eq:mu}
    \frac{1}{2}[S^S(\cA\vert\rho)+S^S(\cB\vert\rho)]
    \ge-\log\bigl(\max_{i,j}\abs{\braket{a_i}{b_j}}\bigr).
  \end{equation}
\end{MU}

In the case of mutually unbiased bases the Maassen--Uffink relation is
maximally strong and thus tight, equality holding for any of the basis states.
(Note that for two arbitrary observables the entropy sum is in general not
minimized by an eigenstate of either of them.~\cite{GhMR03})

\subsection{The stabilizer formalism and graph states}
\label{sec:bases}

As our main tool, we will employ the stabilizer formalism. This formalism
allows to describe certain many-qubit states, among them graph states, in an
efficient manner. For a review of this topic see, e.\,g.,
Ref.~\onlinecite{HDER06}.

The $n$-qubit Pauli group consists of all tensor products of $n$ Pauli
matrices, including the identity, with prefactors $\pm 1$ and $\pm\ir$. Any
commutative subgroup of the Pauli group that has $2^n$ elements and does not
contain $-\one$ has a unique common eigenstate with eigenvalue $+1$. This state
is called stabilizer state; the group is called the stabilizer group and the
group elements are called the stabilizing operators of the state. The
stabilizer group defines in fact a complete basis of common eigenstates, the
stabilizer state being one of them. We refer to this basis as stabilizer basis.
Any basis state is again a stabilizer state, whose stabilizer group is obtained
from the original one by flipping the signs of some of its elements.

\begin{figure}
  \centering
  \includegraphics[width=0.4\textwidth]{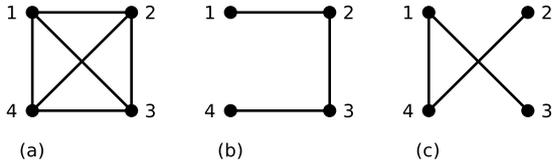}
  \caption{\label{fig:graphs}
    Graphs of the graph states discussed in the text.}
\end{figure}

The most important examples of stabilizer states are graph states. In fact, it
can be shown that any stabilizer state is equivalent to a graph state under a
local unitary operation (more specifically, a local Clifford
operation).~\cite{HDER06} A graph state is described by a simple undirected
graph, whose vertices represent qubits and whose edges represent the
interactions that have created the graph state from a product state [see
Fig.~\ref{fig:graphs} for examples].
More precisely, with the completely disconnected $n$-vertex graph, we associate
the state $\ket{+}^{\otimes n}$, where $\ket{+}$ is the eigenvector of
$\sigma_x$ with eigenvalue $+1$. An edge between two vertices stands for the
controlled phase gate applied to the corresponding pair of qubits, which in the
standard $\sigma_z$-basis is $C=\diag(1,1,1,-1)$. The stabilizing operators of
a graph state can immediately be read off the graph: with qubit $i$, we
associate the operator
\begin{equation}
  K_i=\sigma_x^{(i)}\prod_{j\in N(i)}\sigma_z^{(j)},
\end{equation}
the product being over the neighbourhood $N(i)$ of vertex $i$, that is, all
vertices directly connected to it by an edge. Here $\sigma_x^{(i)}$ and
$\sigma_z^{(i)}$ denote the Pauli matrices $\sigma_x$ and $\sigma_z$ acting on
qubit $i$. The operators $K_i$ generate the stabilizer group. Any state of the
stabilizer basis (here called graph state basis) can be obtained from the graph
state by applying the operation $\sigma_z$ on a subset of the qubits.

As an example, let us consider the graphs in Fig.~\ref{fig:graphs} (a) and (b).
The generators of the stabilizer group of the first state are
$K_1=\sigma_x\otimes\sigma_z\otimes\sigma_z\otimes\sigma_z$,
$K_2=\sigma_z\otimes\sigma_x\otimes\sigma_z\otimes\sigma_z$,
$K_3=\sigma_z\otimes\sigma_z\otimes\sigma_x\otimes\sigma_z$,
and
$K_4=\sigma_z\otimes\sigma_z\otimes\sigma_z\otimes\sigma_x$;
for the second state, they are
$K_1=\sigma_x\otimes\sigma_z\otimes\one\otimes\one$,
$K_2=\sigma_z\otimes\sigma_x\otimes\sigma_z\otimes\one$,
$K_3=\one\otimes\sigma_z\otimes\sigma_x\otimes\sigma_z$,
and
$K_4=\one\otimes\one\otimes\sigma_z\otimes\sigma_x$.
Under local unitary operations, the corresponding graph states are equivalent
to the 4-qubit Greenberger--Horne--Zeilinger (GHZ) state
$\ket{\text{GHZ}_4}=(\ket{0000}+\ket{1111})/\sqrt{2}$
and the 4-qubit linear cluster state
$\ket{\text{C}_4}=(\ket{0000}+\ket{0011}+\ket{1100}-\ket{1111})/2$,
respectively.

\section{A strong uncertainty relation for stabilizer bases}
\label{sec:strong}

In this section, we investigate the conditions under which the Maassen--Uffink
uncertainty relation is tight. We show that this is the case for the
measurement in any two stabilizer bases.

We shall need the following lemma:

\begin{lemma}
  \label{lem:tight}
  If a pair of bases $\cA=\{\ket{a_i}\}$ and $\cB=\{\ket{b_i}\}$ satisfies
  \begin{equation}
    \bigabs{\braket{a_i}{b_j}}\in\{0,r\}\quad\forall\,i,j
  \end{equation}
  for some $r$, then the Maassen--Uffink relation Eq.~\eqref{eq:mu} for the
  measurement in these bases is tight. Equality occurs for any of the basis
  states.
\end{lemma}

\begin{proof}
  For $\rho=\ketbra{b_{j_0}}$ the Maassen--Uffink relation reads
  \begin{equation}
    -\sum_ip_i\log(p_i)\ge-\log(\max_ip_i)
      \quad \text{where}\
    p_i=\bigabs{\braket{a_i}{b_{j_0}}}^2.
  \end{equation}
  Note that the right-hand side is the min-entropy of the probability
  distribution $p_i$. By assumption $p_i\in\{0,r^2\}$ for all $i$. It follows
  that equality holds.
\end{proof}

The main result of this section is the following theorem:

\begin{theorem}
  \label{thm:mustab}
  For the measurement in a pair of stabilizer bases, the Maassen--Uffink
  uncertainty relation Eq.~\eqref{eq:mu} is tight. The bound is attained by any
  of the basis states.
\end{theorem}

The proof is based on a result on mutually unbiased bases, which is due to
Bandyopadhyay \emph{et al.} (see the proof of Theorem~3.2 in
Ref.~\onlinecite{BBRV02}):

\begin{theorem}[Ref.~\onlinecite{BBRV02}]
  Let $C_1$ and $C_2$ each be a set of $d$ commuting unitary $d\times
  d$-matrices. Furthermore assume that $C_1\cap C_2=\{\one\}$ and that all
  matrices in $C_1\cup C_2$ are pairwise orthogonal with respect to the
  Hilbert--Schmidt scalar product. Then the eigenbases defined by either set of
  matrices are mutually unbiased.
\end{theorem}

\begin{proof}[Proof of Theorem~\ref{thm:mustab}]
  Throughout the proof we consider two stabilizing operators that differ only
  by a minus sign as equal. Let $S_1$ and $S_2$ be the two stabilizer groups
  and define $C_0=S_1\cap S_2$. Then $C_0$ is a subgroup of both $S_1$ and
  $S_2$. We consider the factor groups $C_1=S_1/C_0$ and $C_2=S_2/C_0$. The
  groups $C_0$, $C_1$, and $C_2$ are all stabilizer groups, though the spaces
  stabilized by them are in general not one-dimensional. This gives us a
  decomposition of the Hilbert space $\cH=\cH_0\otimes\cH_{12}$, where $C_0$
  acts trivially on $\cH_{12}$ and $C_1$ and $C_2$ act trivially on $\cH_0$. By
  the previous theorem, $C_1$ and $C_2$ define mutually unbiased bases
  $\ket{c^{(1)}_i}$ and $\ket{c^{(2)}_i}$ of $\cH_{12}$. It follows that the
  stabilizer bases can be written as
  $\ket{s^{(1)}_{ij}}=\ket{c^{(0)}_i}\otimes\ket{c^{(1)}_j}$ and
  $\ket{s^{(2)}_{ij}}=\ket{c^{(0)}_i}\otimes\ket{c^{(2)}_j}$, respectively,
  where $\ket{c^{(0)}_i}$ is the basis of $\cH_0$ defined by $C_0$. The
  stabilizer bases thus satisfy the condition of Lemma~\ref{lem:tight} with
  $r=(\dim H_{12})^{-1/2}$.
\end{proof}

In Appendix~\ref{app:altproof}, we give an alternative proof that does not
require the result on mutually unbiased bases. In
Appendix~\ref{app:recurrence}, we develop a method to calculate the right-hand
side of the uncertainty relation explicitly for certain classes of graph
states.

\section{Uncertainty relations for several dichotomic anticommuting
observables}
\label{sec:anticommuting}

Little is known about uncertainty relations for more than two
measurements~\cite{WeWi10} (see, however, Ref.~\onlinecite{MaWB10}). Following
Wehner and Winter,~\cite{WeWi08} we will concentrate on dichotomic
anticommuting observables. An observable is called dichotomic if it has exactly
two distinct eigenvalues. We will always normalize dichotomic observables such
that their eigenvalues are $\pm 1$. In other words, these observables square to
the identity.

The following result has been called a meta-uncertainty
relation,~\cite{WeWi08,WeWi10} for reasons that soon will become apparent.

\begin{lemma}
  \label{lem:anticomm}
  Let $A_1,\ldots,A_L$ be observables which anticommute pairwise
  $\anticomm{A_k}{A_\ell}=0$ for $k\neq\ell$ and which have eigenvalues $\pm
  1$. Then $\sum_{k=1}^L\ev{A_k}^2\le 1$, or equivalently,
  \begin{equation}
    \sum_{k=1}^L\Delta^2(A_k)\ge L-1,
  \end{equation}
  where $\Delta^2(A)=\ev{(A-\ev{A})^2}$ is the variance of $A$.
\end{lemma}

The following proof of this lemma was given in Ref.~\onlinecite{ToGu05}. For an
alternative proof, based on the Clifford algebra, see Ref.~\onlinecite{WeWi08}.

\begin{proof}
  Choose real coefficients $\lambda_1,\ldots,\lambda_L$ with
  $\sum_{k=1}^L\lambda_k^2=1$. Because of anticommutativity and $A_k^2=\one$,
  we have $(\sum_{k=1}^L\lambda_kA_k)^2=\sum_{k=1}^L\lambda_k^2A_k^2
  =\sum_{k=1}^L\lambda_k^2\one=\one$ and thus
  $\abs{\sum_{k=1}^L\lambda_k\ev{A_k}}=\abs{\ev{\sum_{k=1}^L\lambda_kA_k}}\le
  1$ for all states, since for any observable $\ev{X}^2\le\ev{X^2}$.
  Interpreting the expression $\sum_{k=1}^L\lambda_k\ev{A_k}$ as the euclidian
  scalar product of the vector of coefficients $\lambda_k$ and the vector of
  expectation values $\ev{A_k}$, and noting that the vector of coefficients
  $\lambda_k$ is an arbitrary unit vector, we see that the vector $\ev{A_k}$
  has a length less than or equal to $1$. Observing $\sum_{k=1}^L\ev{A_k^2}=L$,
  we obtain the lemma.
\end{proof}

The converse implication is also true in the following sense, as was already
shown in Ref.~\onlinecite{WeWi08}:

\begin{lemma}
  \label{lem:reverse}
  Let $A_1,\ldots,A_L$ be dichotomic anticommuting observables as above, and
  let $a_1,\ldots,a_L$ be real numbers with $\sum_{k=1}^La_k^2\le 1$. Then
  there exists a quantum state $\rho$ such that the numbers $a_k$ are the
  expectation values of the observables, $a_k=\tr(A_k\rho)$.
\end{lemma}

\begin{proof}
  Consider the state $\rho=\frac{1}{d}(\one+\sum_{k=1}^La_kA_k)$, where $d$ is
  the dimension of the Hilbert space. Because of the properties of the
  observables, $\tr(A_kA_\ell)=d\delta_{k\ell}$. Furthermore, the observables
  $A_k$ are traceless: $\tr(A_k)=\tr(A_kA_\ell A_\ell)=\tr(A_\ell
  A_kA_\ell)=-\tr(A_kA_\ell A_\ell)=-\tr(A_k)$. This shows that the state
  $\rho$ has the desired expectation values. It remains to show that $\rho\ge
  0$. But in the proof of the previous lemma, we have already seen that
  $\abs{\sum_{k=1}^La_k\ev{A_k}}\le 1$.
\end{proof}

The meta-uncertainty relation is thus the best possible bound on the
expectation values of the observables. Note that in the case of one qubit and
the three Pauli matrices, it reduces to the Bloch sphere picture. The relation
has also been used to study monogamy relations for Bell
inequalities.~\cite{KPRL11} Generalizing Wehner and Winter's result for the
Shannon entropy,~\cite{WeWi08} we can derive entropic uncertainty relations for
various entropies from it. 

Let $A$ be an observable with eigenvalues $\pm 1$ and $x=[\tr(A\rho)]^2$ its
squared expectation value. Then the probability distribution for the
measurement outcomes of $A$ is given by
$P=(\frac{1+\sqrt{x}}{2},\frac{1-\sqrt{x}}{2})$ or
$P=(\frac{1-\sqrt{x}}{2},\frac{1+\sqrt{x}}{2})$. Any entropy $S$, being
invariant under permutation of $P$, is thus a function of $x$, which we denote
by $\widetilde{S}$,
\begin{equation}
  \label{eq:tilde}
  \widetilde{S}(x)
  =S(A\vert\rho)
  =S\bigl((\frac{1\pm\sqrt{x}}{2},\frac{1\mp\sqrt{x}}{2})\bigr).
\end{equation}
We say that the entropy $S$ is concave in the squared expectation value if the
function $\widetilde{S}$ is concave. This property is the crucial condition for
the following entropic uncertainty relation. We shall also assume that for the
peaked probability distribution, the entropy has the value zero,
$\widetilde{S}(1)=0$.

\begin{theorem}
  \label{thm:anticomm}
  Let $A_1,\ldots,A_L$ be observables which anticommute pairwise
  $\anticomm{A_k}{A_\ell}=0$ for $k\neq\ell$ and which have eigenvalues $\pm 1$
  and let $S$ be an entropy which is concave in the squared expectation value
  (that is, an entropy for which the function $\widetilde{S}$ defined in
  Eq.~\eqref{eq:tilde} is concave). Then
  \begin{equation}
    \min_\rho\frac{1}{L}\sum_{k=1}^LS(A_k\vert\rho)
    =\frac{L-1}{L}S_0,
  \end{equation}
  where $S_0=S\bigl((\frac{1}{2},\frac{1}{2})\bigr)$ is the entropy value of
  the flat probability distribution.
\end{theorem}

\begin{proof}
  For the case of the Shannon entropy the proof was given in
  Ref.~\onlinecite{WeWi08}. Let $x_k=[\tr(A_k\rho)]^2$.
  Lemma~\ref{lem:anticomm} states that $\vec{x}$ lies in the simplex defined by
  $\sum_{k=1}^Lx_k\le 1$ and $x_k\ge 0$. As the function $\widetilde{S}$ is
  concave on the interval $[0,1]$, the function
  $\vec{x}\mapsto\sum_k\widetilde{S}(x_k)$ is concave on the simplex. Thus it
  attains its minimum at an extremal point of the simplex, that is, $x_k=1$ for
  one $k$ and $x_\ell=0$ for $\ell\neq k$. At an extremal point,
  $1/L\sum_{k=1}^L\widetilde{S}(x_k)=S_0(L-1)/L$.
\end{proof}

Before commenting on the implications of this theorem, we discuss which
entropies satisfy the requirement of being concave in the squared expectation
value. For the Shannon entropy, this was already shown in
Ref.~\onlinecite{WeWi08}. The Tsallis entropy can be treated analogously,
though one has to distinguish between different parameter ranges.

\begin{lemma}
  \label{lem:tsallis}
  The Tsallis entropy $S_q^T$ of a dichotomic observable is concave in the
  squared expectation value (that is, the function $\widetilde{S}_q^T$ defined
  as in Eq.~\eqref{eq:tilde} is concave on the interval $[0,1]$) for parameter
  values $1<q<2$ and $3<q$, but convex for $2<q<3$.
\end{lemma}

\begin{proof}
  Explicitly,
  \begin{equation}
    \widetilde{S}_q^T(x)
    =\frac{1}{q-1}
    \Bigl[1-\Bigl(\frac{1+\sqrt{x}}{2}\Bigr)^q
    -\Bigl(\frac{1-\sqrt{x}}{2}\Bigr)^q\,\Bigr].
  \end{equation}
  For $q=2$ and $q=3$, this function is easily seen to be linear. For the
  second derivative, we obtain
  \begin{multline}
    \partial_x^2\widetilde{S}_q^T(x)=
    \frac{q}{q-1}\frac{1}{2^{q+2}}\frac{1}{x^{3/2}}
    \Bigl\{
    (1+\sqrt{x})^{q-2}\bigl[1-\sqrt{x}(q-2)\bigr]\\
    -(1-\sqrt{x})^{q-2}\bigl[1+\sqrt{x}(q-2)\bigr]
    \Bigr\}.
  \end{multline}
  Substituting $y=\sqrt{x}$ and omitting the prefactor (which is always
  positive), we arrive at the function
  \begin{equation}
    f_q(y)=(1+y)^{q-2}[1-y(q-2)]-(1-y)^{q-2}[1+y(q-2)].
  \end{equation}
  Observing that $f_q(0)=0$, we note that $f_q(y)$ is positive (negative) for
  all $0<y\le 1$ if its derivative $f'_q(y)$ is positive (negative) for all
  $0<y\le 1$. The derivative is given by
  \begin{equation}
    f'_q(y)=-(q-2)(q-1)y\bigl[(1+y)^{q-3}-(1-y)^{q-3}\bigr].
  \end{equation}
  For $1<q<2$, the prefactor $-(q-2)(q-1)$ is positive and the term in the
  square brackets is negative; for $2<q<3$, the prefactor is negative and the
  term in the brackets is still negative; for $q>3$, the prefactor is negative
  and the term in the brackets positive. This proves the lemma.
\end{proof}

Let us add some remarks on the theorem. The Shannon entropy has the required
concavity in the squared expectation value, and the resulting uncertainty
relation is the one found by Wehner and Winter.~\cite{WeWi08} For the Tsallis
entropy $S_q^T$, we have to distinguish between different parameter ranges: for
parameter values $q=2$ and $q=3$ this entropy is, up to a constant factor,
equal to the variance, $S_2^T(A\vert\rho)=1/2\Delta^2(A)$ and
$S_3^T(A\vert\rho)=3/8\Delta^2(A)$, and the uncertainty relation is equivalent
to the meta-uncertainty relation itself. Thus it is the optimal uncertainty
relation for these observables. The relation based on the Shannon entropy is
strictly weaker.

In Lemma~\ref{lem:tsallis}, we have shown that the Tsallis entropy satisfies
the condition of Theorem~\ref{thm:anticomm} for parameter values $1<q\le 2$ and
$3\le q$. The entropy value for the flat probability distribution, which
determines the bound, is $S_0=(1-2^{1-q})/(q-1)$. In the special case of the
observables $\sigma_x$ and $\sigma_y$ and parameter $q\in[2n-1,2n]$ with
$n\in\N$, the uncertainty relation was derived before in Footnote~32 of
Ref.~\onlinecite{GuLe04}.

\begin{figure}
  \centering
  \includegraphics[width=0.45\textwidth]{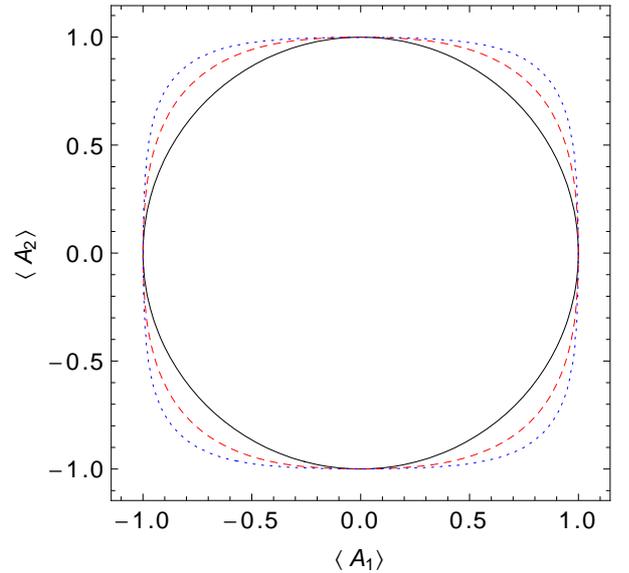}
  \caption{\label{fig:circles}
    Bounds on the expectation values $(\ev{A_1},\ev{A_2})$ for two dichotomic
    anticommuting observables provided by different uncertainty relations. The
    black solid line corresponds to the meta-uncertainty relation
    Lemma~\ref{lem:anticomm}, which can also be understood as an entropic
    uncertainty relation for the Tsallis entropy with parameter value $q=2$ or
    $q=3$. The red dashed line and the blue dotted line correspond to the
    entropic uncertainty relation Theorem~\ref{thm:anticomm} for the Shannon
    entropy and the Tsallis entropy with $q=8$, respectively.}
\end{figure}

As we remarked above, Lemmas~\ref{lem:anticomm} and~\ref{lem:reverse} provide a
complete characterization of the set of expectation values of dichotomic
anticommuting observables which can originate from valid quantum states.
Deriving uncertainty relations from them means approximating this set from the
outside. This is illustrated in Fig.~\ref{fig:circles}. 

In the parameter range $2<q<3$, the Tsallis entropy does not satisfy the
condition for the theorem (see Lemma~\ref{lem:tsallis}). An exceptional
behaviour of the Tsallis entropy in this parameter range was also reported in
Ref.~\onlinecite{GuLe04}. The collision entropy or R\'enyi entropy of order $2$
and the min-entropy do not satisfy the condition either. The uncertainty
relations for these entropies given in Ref.~\onlinecite{WeWi08} also follow
from the meta-uncertainty relation, but do not fit into this scheme.

In the following section, we will apply Theorem~\ref{thm:anticomm} to the
stabilizing operators of two stabilizer states.

Uncertainty relations for several observables can also be constructed without
requiring anticommutativity: applying a result by Mandayam \emph{et
al.}~\cite{MaWB10} to a set of stabilizer bases $\cA_1,\ldots,\cA_L$ with basis
vectors denoted by $\cA_k=\{\ket{a^{(k)}_i}\}_i$, one finds for their
min-entropies
\begin{equation}
\frac{1}{L}\sum_{k=1}^LS^\text{min}(\cA_k\vert\rho)
\ge-\log\Bigl[\frac{1+r(L-1)}{L}\Bigr],
\end{equation}
where
\begin{equation}
r=\max_{k\neq\ell}\max_{i,j}\bigabs{\braket{a_i^{(k)}}{a_j^{(\ell)}}}
\end{equation}
is the maximal overlap of the basis states. We omit the proof, since this
relation readily follows from Appendix~C.1 of Ref.~\onlinecite{MaWB10}.

\section{An uncertainty relation for stabilizing operators}
\label{sec:stabilizer}

In this section, we will apply the uncertainty relation for anticommuting
dichotomic observables (Theorem~\ref{thm:anticomm}) to the elements of a pair
of stabilizer groups.

Let $S$ and $T$ be two $n$-qubit stabilizer groups. Throughout this section, we
consider two operators as equal if they differ only by a minus sign. Define
$\widetilde{S}=S\setminus T$ and $\widetilde{T}=T\setminus S$. Let
$M=\widetilde{S}\cup\widetilde{T}$ be the symmetric difference of $S$ and $T$
and denote its elements by $A_1,\ldots,A_L$. In Theorem~\ref{thm:groups}, we
will give a lower bound on $1/L\sum_{k=1}^LS(A_k\vert\rho)$.

We begin by proving the following lemma.

\begin{lemma}
  Let $S$ be a stabilizer group and $g$ a Pauli operator (that is, a tensor
  product of Pauli matrices and the identity matrix) which anticommutes with an
  element $s_0\in S$. Then $g$ anticommutes with exactly half of the elements
  of $S$.
\end{lemma}

\begin{proof}
  Choose any $s_1\in S$ with $s_1\neq s_0$ and let $s_2=s_0s_1$. We will now
  show that $g$ anticommutes with $s_2$ if it commutes with $s_1$ and vice
  versa. Consider first the case $\comm{s_1}{g}=0$. The identity
  $\anticomm{AB}{C}=A\comm{B}{C}+\anticomm{A}{C}B$ shows that
  $\anticomm{s_2}{g}
  =\anticomm{s_0s_1}{g}=s_0\comm{s_1}{g}+\anticomm{s_0}{g}s_1=0$. Consider now
  the case $\anticomm{s_1}{g}=0$. Using the same identity, we obtain
  $s_0\comm{s_2}{g}=\anticomm{s_0s_2}{g}-\anticomm{s_0}{g}s_2
  =\anticomm{s_1}{g}-\anticomm{s_0}{g}s_2=0$ and thus $\comm{s_2}{g}=0$. We now
  iterate this procedure by choosing $s_3\in S\setminus\{s_0,s_1,s_2\}$ and
  using it in place of $s_1$. (Note that $s_0s_3\neq s_1$ and $\neq s_2$.) The
  operator $s_0$ is kept fixed during the whole iteration. In this way, $S$ can
  be divided into pairs of observables, each consisting of one element
  commuting with $g$ and one anticommuting with $g$. Note that $s_0$ forms a
  pair with the identity.
\end{proof}

Excluding the trivial case $S=T$, any element of $T$ anticommutes with at least
one element of $S$, since at most $2^n$ orthogonal (with respect to the
Hilbert--Schmidt scalar product) unitary $2^n\times 2^n$-matrices can commute
pairwise (see, e.\,g., Ref.~\onlinecite{BBRV02}). Thus $L=\abs{M}$ varies from
$2^n$ to $2(2^n-1)$.
 
We will now see, using only combinatorical reasoning, that this lemma implies
that $M$ can be divided into anticommuting pairs. We shall need the following
combinatorical result:

\begin{marriage}
  Consider a bipartite graph, that is, two disjoint sets of vertices $U$ and
  $V$ and a collection of edges, each connecting a vertex in $U$ with a vertex
  in $V$. We consider the case of $\abs{U}=\abs{V}$. Then the graph contains a
  perfect matching, that is, the vertices can be divided into disjoint pairs of
  connected vertices, if and only if the following marriage condition is
  fulfilled: For each subset $U'$ of $U$, the set $V'$ of vertices in $V$
  connected to vertices in $U'$ is at least as large as $U'$.
\end{marriage}

This theorem was first proven in Ref.~\onlinecite{Hall35}.

\begin{lemma}
  \label{lem:matching}
  The symmetric difference $M$ of any two stabilizer groups $S$ and $T$ can be
  divided into anticommuting pairs of operators.
\end{lemma}

\begin{proof}
  We show that the marriage condition is fulfilled. Let $S'$ be any subset of
  $\widetilde{S}$. Consider first the case $\abs{S'}>2^{n-1}$. Then any $t\in
  T$ anticommutes with at least one $s\in S'$, because any such $t$
  anticommutes with exactly $2^{n-1}$ elements of $\widetilde{S}$. Thus the
  number of $t\in\widetilde{T}$ anticommuting with at least one $s\in S'$ is
  $\abs{\widetilde{T}}=\abs{\widetilde{S}}\ge\abs{S'}$. Consider now the case
  $\abs{S'}\le 2^{n-1}$. For any $s\in S'$, we then find $2^{n-1}$ elements of
  $\widetilde{T}$ anticommuting with $s$. Thus the number of
  $t\in\widetilde{T}$ anticommuting with at least one $s\in S'$ is
  $2^{n-1}\ge\abs{S'}$.
\end{proof}

We are now ready to state the main result of this section.

\begin{theorem}
  \label{thm:groups}
  Let $M=\{A_1,\ldots,A_L\}$ be the symmetric difference of two stabilizer
  groups. Then
  \begin{equation}
    \frac{1}L\sum_{k=1}^LS(A_k\vert\rho)
    \ge\frac{1}{2}S_0
  \end{equation}
  holds, where $S$ is an entropy which is concave in the squared expectation
  value (that is, an entropy for which the function $\widetilde{S}$ defined in
  Eq.~\eqref{eq:tilde} is concave) and $S_0$ is the entropy value of the flat
  probability distribution. Any basis state of either stabilizer basis attains
  the lower bound.
\end{theorem}

\begin{proof}
  Due to Theorem~\ref{thm:anticomm} the uncertainty relation
  $S(A_k\vert\rho)+S(A_\ell\vert\rho)\ge S_0$ holds for any anticommuting pair
  $A_k$, $A_\ell$. Lemma~\ref{lem:matching} states that $M$ consists of $L/2$
  such pairs. This shows the uncertainty relation. The density matrix of the
  stabilizer state defined by the group $T=\{T_k\}$ is given by
  \begin{equation}
    \rho_T=\frac{1}{2^n}\sum_{k=1}^{2^n}T_k
  \end{equation}
  (see, e.\,g., Ref.~\onlinecite{HDER06}). Thus
  \begin{equation}
    \tr(A_k\rho_T)=\frac{1}{2^n}\sum_{\ell=1}^{2^n}\tr(A_kT_\ell)=0
    \quad\text{for all}\quad A_k\notin T,
  \end{equation}
  showing $S(A_k\vert\rho_T)=S_0$ for $L/2$ observables $A_k$.
\end{proof}

This relation is not maximally strong. This is due to the fact that some of the
observables commute.

\section{Conclusion}
\label{sec:conclusion}

We demonstrated how the stabilizer formalism can be combined with the theory of
entropic uncertainty relations. We focussed on two problems: the
characterization of pairs of measurement bases which give rise to strong
uncertainty relations and the generalization of the theory to the
many-observable setting.

Concerning the first question, we showed that for the measurement in any two
stabilizer bases the Maassen--Uffink relation is tight
(Theorem~\ref{thm:mustab}). We  also demonstrated how the stabilizer formalism
can be used to compute the overlap of the basis states, which gives the lower
bound on the entropy sum.

Concerning the second question, we generalized a result by Wehner and Winter on
the Shannon entropy for several dichotomic anticommuting observables to a
larger class of entropies (Theorem~\ref{thm:anticomm}). Comparing the strengths
of these uncertainty relation, we saw that entropic relations are not
necessarily stronger than variance-based ones. Indeed, in the case of
dichotomic anticommuting observables the variance-based uncertainty relation is
optimal in the sense that it exactly describes the set of expectation values
which can originate from valid quantum states. As an application of
Theorem~\ref{thm:anticomm}, we derived an uncertainty relation for the elements
of two stabilizer groups (Theorem~\ref{thm:groups}). 

\begin{acknowledgments}
  We thank Bastian Jungnitsch, Oleg Gittsovich, Tobias Moroder, and Gemma De
  las Cuevas for helpful discussions, and Maarten Van den Nest for pointing out
  Refs. \onlinecite{DeDM03} and \onlinecite{VdNe10}. Furthermore, we thank the
  associate editor and an anonymous referee for helpful comments, in particular
  for a simple proof of Lemma 7. This work has been supported by the Austrian
  Science Fund (FWF): Y376-N16 (START prize) and the EU (Marie Curie CIG
  293993/ENFOQI). 
\end{acknowledgments}

\appendix

\section{Alternative proof of Theorem~\ref{thm:mustab}}
\label{app:altproof}

In this appendix, we give an alternative proof of Theorem~\ref{thm:mustab},
which is not based on any previous results on mutually unbiased bases. Let
$S=\{S_k\}$ and $T=\{T_k\}$ be two $n$-qubit stabilizer groups with stabilizer
states $\ket{S}$ and $\ket{T}$. Define $S^+=S\cap T$, where, unlike in the
first proof, we consider two operators as distinct if they differ by a minus
sign. Also define $S^-=S\cap-T$. Both $S^+$ and $S^+\cup S^-$ are easily seen
to be subgroups of $S$. By Lagrange's theorem $\abs{S^+}=2^p$ and $\abs{S^+\cup
S^-}=2^q$ with some $p\in\{1,2,\ldots,n\}$ and $q\in\{p,p+1,\ldots,n\}$. The
projectors onto the stabilizer states are given by
\begin{equation}
  \ketbra{S}=\frac{1}{2^n}\sum_{k=1}^{2^n}S_k
\end{equation}
and similarly for $\ket{T}$ (see, e.\,g., Ref.~\onlinecite{HDER06}). Thus
\begin{equation}
  \begin{split}
    \bigabs{\braket{S}{T}}^2
    &=\frac{1}{2^{2n}}\sum_{k,\ell=1}^{2^n}\tr(S_kT_\ell)\\
    &=\frac{1}{2^n}\bigl(\abs{S^+}-\abs{S^-}\bigr)\\
    &=\frac{1}{2^n}(2^{p+1}-2^q)\\
    &=
    \begin{cases}
      2^{q-n} & \text{for $p=q$,}\\
      0 & \text{for $p=q-1$.}
    \end{cases}
  \end{split}
\end{equation}
The case $p<q-1$ cannot occur since it would give a negative value of
$\abs{\braket{S}{T}}^2$ and thus lead to a contradiction.

Consider now another state $\ket{T'}$ of the stabilizer basis of $T$. This
state is again a stabilizer state, whose stabilizing operators are equal to
those of $\ket{T}$ up to some minus signs. In particular $S^+\cup S^-$ and thus
$q$ are the same for $\ket{T}$ and for $\ket{T'}$, and Lemma~\ref{lem:tight}
applies.\hfill\qedsymbol

\section{A recurrence relation for the overlap of two graph state bases}
\label{app:recurrence}

In this appendix, we derive a recurrence relation for the scalar products
$\braket{G^{(1)}_i}{G^{(2)}_j}$ of two graph state bases $\ket{G^{(1)}_i}$ and
$\ket{G^{(2)}_i}$ and use it to determine the Maassen--Uffink bound for certain
classes of graph states. Recall that any of these basis states is obtained from
the state $\ket{+}^{\otimes n}$ by applying first controlled phase gates
$C=\diag(1,1,1,-1)$ and then local phases $\sigma_z$. Since all these
operations commute, we can move all phase gates in the scalar product
$\braket{G^{(1)}_i}{G^{(2)}_j}$ to the right and all local phases to the left.
This corresponds to replacing $G^{(1)}$ by the empty or completely disconnected
graph and $G^{(2)}$ by the ``sum modulo $2$'' of the graphs $G^{(1)}$ and
$G^{(2)}$. Similarly, it does not restrict the set of values of
$\braket{G^{(1)}_i}{G^{(2)}_j}$ if we consider only one state of the basis
$\ket{G^{(2)}_j}$, such as the graph state $\ket{G^{(2)}}$ itself. The graph
state basis of the empty graph consists of all tensor products of the
eigenstates of $\sigma_x$. We can write those states as $H^{\otimes
n}\ket{\vec{y}}$, where
$H=\frac{1}{\sqrt{2}}\bigl(
\begin{smallmatrix}
1 & 1 \\
1 & -1
\end{smallmatrix}
\bigr)$
is the Hadamard gate and $\ket{\vec{y}}$ for $\vec{y}\in\F_2^n$ is a state of
the standard basis in binary notation.

[As the Hadamard gate is a local Clifford operation, the state $H^{\otimes
n}\ket{G^{(2)}}$ is a stabilizer state. This shows that the scalar products
$\braket{G^{(1)}_i}{G^{(2)}}$ can be understood as the coefficients of a
stabilizer state with respect to the standard basis. It has been shown that for
any stabilizer state these coefficients are $0$, $\pm 1$, and $\pm\ir$, up to a
global normalization (see Theorem~5 and the paragraph below in
Ref.~\onlinecite{DeDM03}. For an alternative proof, see
Ref.~\onlinecite{VdNe10}.) This constitutes yet another proof of
Theorem~\ref{thm:mustab} for the case of graph state bases.]

As an example, consider the graphs in Fig.~\ref{fig:graphs} (a) and (b). Up to
local unitaries, the corresponding graph states are the 4-qubit GHZ state and
the 4-qubit linear cluster state, but the uncertainty relation is not invariant
under these local unitary operations. By the above remark the Maassen--Uffink
bound for the corresponding bases is equal to the bound for the empty graph and
the ``sum'' of the graphs, which in our case is given by
Fig.~\ref{fig:graphs}~(c). The latter is again equal to the graph~(b), up to a
permutation of vertices.

Let us return to the explicit calculation of the overlaps. In the standard
basis, we have
\begin{equation}
  H^{\otimes n}\ket{\vec{y}}
  =\frac{1}{2^{n/2}}\sum_{\vec{x}\in\F_2^n}(-1)^{\sum_iy_i x_i}\ket{\vec{x}}.
\end{equation}
The adjacency matrix of an $n$-qubit graph is the symmetric $n\times n$-matrix
$A$ whose entry $A_{ij}$ is $1$ if there is an edge between vertices $i$ and
$j$ and $0$ if there is not. It thus provides a complete description of the
graph state, which we will therefore denote by $\ket{G_A}$. The representation
of this state in the standard basis is (see Proposition~2.14 in
Ref.~\onlinecite{VdNe05})
\begin{equation}
  \ket{G_A}=\frac{1}{2^{n/2}}
  \sum_{\vec{x}\in\F_2^n}(-1)^{\sum_{i<j}x_iA_{ij}x_j}\ket{\vec{x}}.
\end{equation}
The scalar products are thus given by
\begin{equation}
  \begin{split}
    R_n(\vec{y},A)&\coloneqq\bra{\vec{y}}H^{\otimes n}\ket{G_A}\\
    &=\frac{1}{2^n}\sum_{\vec{x}\in\F_2^n}
    (-1)^{\sum_iy_ix_i+\sum_{i<j}x_iA_{ij}x_j}.
  \end{split}
\end{equation}
To derive a recurrence relation, we write
\begin{equation}
  \vec{x}=
  \begin{pmatrix}
    \xi\\
    \vec{x}\,{}'
  \end{pmatrix},
  \quad
  \vec{y}=
  \begin{pmatrix}
    \upsilon\\
    \vec{y}\,{}'
  \end{pmatrix},
  \quad
  A=
  \begin{pmatrix}
    0 & \vec{a}{}'^t\\
    \vec{a}{}' & A'
  \end{pmatrix}
\end{equation}
and obtain
\begin{widetext}
\begin{equation}
  \begin{split}
    R_n(\vec{y},A)
    &=\frac{1}{2^n}\sum_{\xi\in\{0,1\}}
    \sum_{\vec{x}\,{}'\in\F_2^{n-1}}
    (-1)^{\upsilon\xi+\sum_iy'_ix'_i
    +\xi\sum_ia'_ix'_i+\sum_{i<j}x'_iA'_{ij}x'_j}\\
    &=\frac{1}{2^n}\sum_{\vec{x}\,{}'\in\F_2^{n-1}}
    (-1)^{\sum_iy'_ix'_i+\sum_{i<j}x'_iA'_{ij}x'_j}
    +\frac{1}{2^n}(-1)^{\upsilon}
    \sum_{\vec{x}\,{}'\in\F_2^{n-1}}
    (-1)^{\sum_i(y'_i+a'_i)x'_i+\sum_{i<j}x'_iA'_{ij}x'_j}\\
    &=\frac{1}{2}R_{n-1}(\vec{y}\,{}',A')
    +(-1)^\upsilon\frac{1}{2}R_{n-1}(\vec{y}\,{}'+\vec{a}\,{}',A').
  \end{split}
\end{equation}
\end{widetext}
This is the desired recurrence relation.

We already know that $R_{n-1}(\vec{y}\,{}',A')\in\{0,\pm r_{A'},\pm\ir
r_{A'}\}$ for all $\vec{y}$ for some $r_{A'}$ Since the coefficients $R$ are
real, $R_{n-1}(\vec{y}\,{}',A')\in\{0,\pm r_{A'}\}$. Thus we have
$R_n(\vec{y},A)\in\{0,\pm\frac{1}{2}r_{A'},\pm r_{A'}\}$. This shows that
either $r_A=\frac{1}{2}r_{A'}$ or $r_A=r_{A'}$, depending on $\vec{a}\,{}'$.

The Maassen--Uffink relation for the graph state bases is maximally strong if
and only if $r=2^{-n/2}$. On the other hand, $r$ is an integer multiple of
$2^{-n}$ (even $2^{1-n}$). One can see this by induction with the recurrence
relation. This shows that this uncertainty relation is never maximally strong
for graph state bases with an odd number of qubits.

We will now use the recurrence relation to compute $r$ for certain classes of
states, and by doing so, construct maximally strong uncertainty relations for
all even numbers of qubits. We assume one graph to be empty and vary only the
other one. The application of the recurrence relation is particularly easy if
$R_{n-1}(\vec{y}\,{}',A')$ is independent of $\vec{y}\,{}'$, up to a sign.

First we show by induction that for the fully connected graph with an even
number $n$ of qubits, we have $R_n(\vec{y},A)=\pm 2^{-n/2}$ for all $\vec{y}$.
For $n=2$, we have $r_2=1/2$. Assume the assertion to be true for $n-2$. Let
$A$, $A'$, and $A''$ be the fully connected adjacency matrices for $n$, $n-1$,
and $n-2$ qubits, respectively, and $\vec{a}\,'=(1,1,\ldots,1)\in\F_2^{n-1}$,
$\vec{a}\,''=(1,1,\ldots,1)\in\F_2^{n-2}$ and similarly for $\vec{0}$. Then
\begin{equation}
  \begin{split}
    R_n(\vec{0},A)&=\frac{1}{2}R_{n-1}(\vec{0}\,{}',A')+
    \frac{1}{2}R_{n-1}(\vec{a}\,{}',A')\\
    &=\frac{1}{2}\Bigl[\frac{1}{2}R_{n-2}(\vec{0}\,{}'',A'')
    +\frac{1}{2}R_{n-2}(\vec{a}\,{}'',A'')\Bigr]\\
    &\qquad
    +\frac{1}{2}\Bigl[\frac{1}{2}R_{n-2}(\vec{a}\,{}'',A'')
    -\frac{1}{2}R_{n-2}(\vec{0}\,{}'',A'')\Bigr]\\
    &=\frac{1}{2}R_{n-2}(\vec{a}\,{}'',A'')\\
    &=\pm\frac{1}{2}\frac{1}{2^{(n-2)/2}}=\pm\frac{1}{2^{n/2}}.
  \end{split}
\end{equation}
This implies that $R_n(\vec{y},A)\in\{0,\pm 2^{-n/2}\}$ for all $\vec{y}$. But
because of normalization, $R_n(\vec{y},A)=0$ is not possible. This shows the
assertion. For the fully connected graph with an odd number of qubits, we have
\begin{equation}
  \begin{split}
    R_n(\vec{y},A)&=\frac{1}{2}R_{n-1}(\vec{y}\,{}',A')
    \pm\frac{1}{2}R_{n-1}(\vec{y}\,{}'+\vec{a}\,{}',A')\\
    &\in\bigl\{0,\pm\frac{1}{2^{(n-1)/2}}\bigr\}.
  \end{split}
\end{equation}

The generalization of the state in Fig.~\ref{fig:graphs}~(b), which is
equivalent under local unitary operations to the linear cluster state, can be
treated in exactly the same way, and we obtain the same results for $r_n$.

\bibliography{entropic}

\end{document}